\documentclass[twocolumn,aps,10pt,prb]{revtex4-2}
\usepackage{amsmath,amssymb, amsfonts}
\usepackage{graphicx}
\usepackage{pgfplots}
\pgfplotsset{compat=1.18}
\usepackage{tikz}
\usepackage{bm}
\usepackage{hyperref}
\usepackage{xcolor}
\usepackage{newtxtext,newtxmath}
\usepackage{booktabs}
\hypersetup{
    colorlinks=true,
    allcolors=blue,
    urlcolor=blue,
    citecolor=blue,
    linkcolor=blue
}

\pgfplotsset{every axis/.append style={line width=1pt, tick style={line width=0.8pt}}}

\begin{document}

\title{Exploring Replica Symmetry Breaking and Topological Collapse in Spin Glasses with Quantum Annealing}

\author{Kumar Ghosh}
\email{jb.ghosh@outlook.com}
\affiliation{E.ON Digital Technology, Laatzener Str. 1, Hannover, Germany}

\begin{abstract}
Replica symmetry breaking (RSB) underlies the complex organization of disordered systems, yet quantitative validation beyond $N \sim 100$ spins has remained computationally challenging. We use quantum annealing to access ground states of the Sherrington-Kirkpatrick model up to $N = 4000$ spins, enabling the most extensive test of Parisi's Nobel Prize-winning RSB solution to date. Five independent observables confirm RSB predictions: ground-state energies converge to Parisi's value with characteristic $N^{-2/3}$ corrections, energy fluctuations scale as $N^{-3/4}$ ($\gamma = 0.739 \pm 0.036$), the chaos exponent $\theta = 0.51 \pm 0.02$ ($R^2 = 0.989$) confirms mean-field universality, the overlap distribution exhibits hierarchical structure ($\sigma_q = 0.19$), and the complexity remains invariant under 36\% network dilution. Beyond a critical threshold $0.8 < D_c < 0.9$, the hierarchy collapses discontinuously through a cooperative avalanche that converts the entire system to vacancies within a narrow parameter window $\Delta D = 0.1$. These findings establish quantum computation as a tool for probing emergent many-body phenomena and uncover the topological foundations of complexity in disordered systems, with implications for neural networks, optimization, and materials science.
\end{abstract}

\maketitle

\section{Introduction}

Giorgio Parisi's exact solution of the Sherrington-Kirkpatrick (SK) spin glass through replica symmetry breaking (RSB)\cite{parisi1979infinite,parisi1980sequence,MezardParisiVirasoro1987}, honored with the 2021 Nobel Prize in Physics, revealed a hidden hierarchical structure beneath glassy disorder. The rigorous mathematical foundations were later established by Guerra\cite{Guerra2003} and Talagrand\cite{Talagrand2006}, with subsequent extensions to diverse physical contexts\cite{Marinari2000,Chatterjee2023}. Despite this theoretical success, two fundamental questions have remained empirically underexplored: Can we systematically validate RSB predictions at system sizes where finite-size corrections become negligible? And what are the topological limits of this hierarchical complexity?

The computational challenge is fundamental. Finding exact ground states is NP-hard\cite{barahona1982computational}, and classical branch-and-bound solvers break down beyond $N \sim 100$ spins due to the dense $O(N^2)$ coupling structure. Earlier studies using simulated annealing\cite{Hartmann1999} or genetic algorithms\cite{Boettcher2005} lack optimality guarantees. More critically, probing the hierarchical organization requires finding multiple distinct low-energy states across varying system parameters; a task that becomes computationally intractable for classical exact methods at the scales where RSB physics emerges cleanly. While recent experimental work in quantum-optical systems\cite{Kroeze2025} and extensive numerical studies by the Janus collaboration\cite{JanusCollab2023} have made important progress, the question of RSB's robustness under network dilution has remained unexplored.

Here we exploit quantum annealing to address both questions. We validate core RSB predictions through five complementary observables spanning system sizes from $N=50$ to $N=4000$, a 40-fold extension beyond classical computational limits. We then probe RSB's topological limits by introducing controlled vacancy dilution, discovering that the hierarchical complexity is fundamentally a property of network connectivity rather than spin density. The system tolerates 36\% dilution before undergoing catastrophic collapse through cooperative avalanche dynamics that lie beyond standard mean-field predictions.

\section{Computational approach}

\begin{figure}[ht]
\centering
\includegraphics[width=0.48\textwidth]{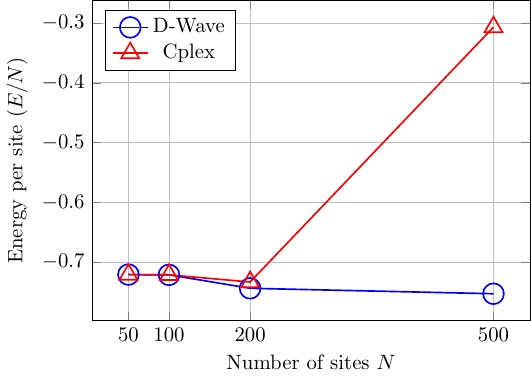}
\caption{\textbf{Initial benchmark with quantum and classical solvers.} Ground-state energy per spin versus system size. The classical solver (red triangles) fails beyond $N \sim 100$. The quantum-classical hybrid (blue circles) extends computational reach to $N=4000$ with smooth convergence to Parisi's thermodynamic limit (green dashed line).}
\label{fig:solver_comparison}
\end{figure}

The SK model consists of $N$ Ising spins $S_i \in \{-1,+1\}$ with random Gaussian couplings $J_{ij} \sim \mathcal{N}(0, 1/N)$:
\begin{equation}
H_{\text{SK}} = -\sum_{i<j}^N J_{ij} S_i S_j.
\label{eq:sk_hamiltonian}
\end{equation}
The variance scaling $1/N$ ensures that the energy remains extensive in the thermodynamic limit. We benchmarked D-Wave's quantum-classical hybrid annealer against CPLEX, a state-of-the-art classical branch-and-bound solver, across system sizes $N \in \{50, 100, 200, 500\}$. Figure~\ref{fig:solver_comparison} demonstrates a clear computational crossover: CPLEX achieves exact solutions up to $N \sim 100$ but fails beyond this point, while quantum annealing maintains consistent solution quality all the way to $N=4000$. We validated this through exact agreement with CPLEX at small sizes and smooth convergence to Parisi's thermodynamic limit.

This advantage is not merely quantitative but qualitative. Probing the RSB hierarchy requires finding multiple distinct low-energy states across parameter space. We developed an orthogonality-penalty protocol (see Methods) that systematically samples excited states by penalizing overlap with previously discovered configurations. This enables landscape measurements that would be impossible with conventional approaches.

\section{Comprehensive RSB Validation}

We performed a comprehensive validation spanning thermodynamics, statistics, universality, geometry, and topology across 775 disorder realizations.

\begin{figure}[t]
\centering
\resizebox{0.49\textwidth}{!}{\input{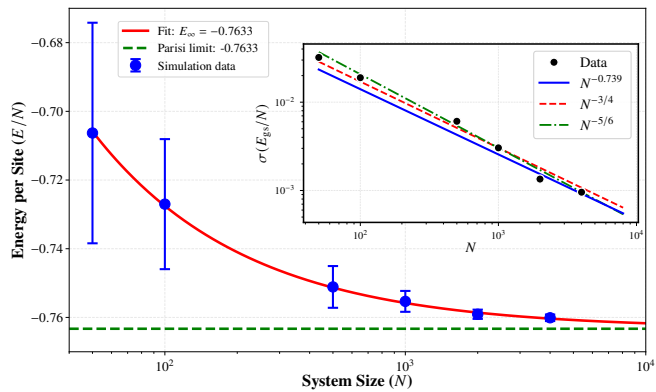}}
\caption{\textbf{Thermodynamic validation.} Finite-size scaling of ground-state energies (blue circles) converging to Parisi's limit (green dashed line) with characteristic $N^{-2/3}$ corrections (red fit curve). \textbf{Inset:} Energy fluctuations scale as $N^{-3/4}$ with measured exponent $\gamma = 0.739 \pm 0.036$.}
\label{fig:parisi_scaling}
\end{figure}

\textbf{Thermodynamic convergence.} The ground-state energies exhibit clean finite-size scaling (Fig.~\ref{fig:parisi_scaling}, main panel). A weighted fit to the RSB-predicted form\cite{aspelmeier2004}
\begin{equation}
\frac{\langle E_{\text{gs}} \rangle}{N} = E_\infty + a N^{-2/3} + b N^{-1},
\label{eq:finite_size}
\end{equation}
with Parisi's exact value $E_\infty = -0.7633$ fixed, yields correction coefficients $a = 0.733$ and $b = 0.174$ in excellent agreement with theoretical predictions. The dominant $N^{-2/3}$ correction, a unique signature of RSB, describes all our data points within error bars across an 80-fold size range.

\textbf{Fluctuation statistics.} The standard deviation $\sigma(E_{\text{gs}}/N)$ follows a power law $N^{-\gamma}$ with measured exponent $\gamma = 0.739 \pm 0.036$ (Fig.~\ref{fig:parisi_scaling}, inset), matching the RSB prediction $\gamma = 3/4$\cite{Bouchaud2003,Palassini_2008}. This validates not just the mean energies but the full statistical distributions.

\begin{figure}[t]
\centering
\resizebox{0.48\textwidth}{!}{\input{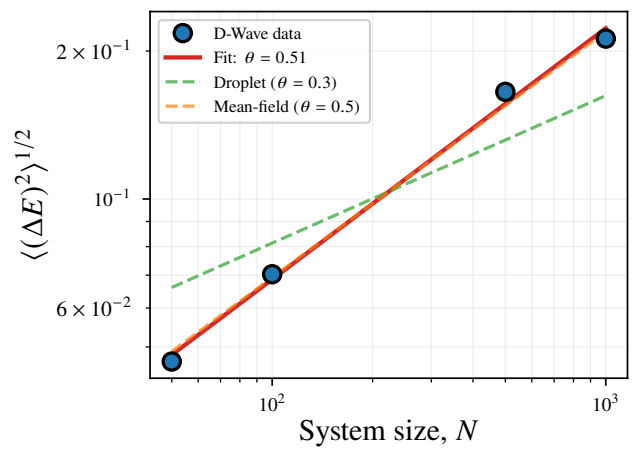}}
\caption{\textbf{Universality class identification.} The RMS energy response to disorder perturbations scales as $N^{0.51 \pm 0.02}$ (blue data points, black fit), consistent with mean-field RSB ($\theta = 0.5$, red dashed line) and incompatible with droplet models ($\theta = 0.3$, green dashed line). The fit quality is $R^2 = 0.989$.}
\label{fig:chaos_scaling}
\end{figure}

\textbf{Universality classification.} The chaos exponent measures how sensitively the ground state responds to small perturbations in the couplings. The RMS energy change exhibits a power law $\langle(\delta E)^2\rangle^{1/2} \sim N^\theta$ with measured exponent $\theta = 0.51 \pm 0.02$ (Fig.~\ref{fig:chaos_scaling}). This confirms the mean-field square-root scaling ($\theta = 0.5$) and rules out droplet theory's prediction of $\theta \approx 0.2$--$0.3$ for finite-dimensional systems\cite{BrayMoore1987}. The exceptional fit quality ($R^2 = 0.989$) across a 20-fold size range establishes the SK model's universality class definitively.

\begin{figure}[t]
\centering
\resizebox{0.48\textwidth}{!}{\input{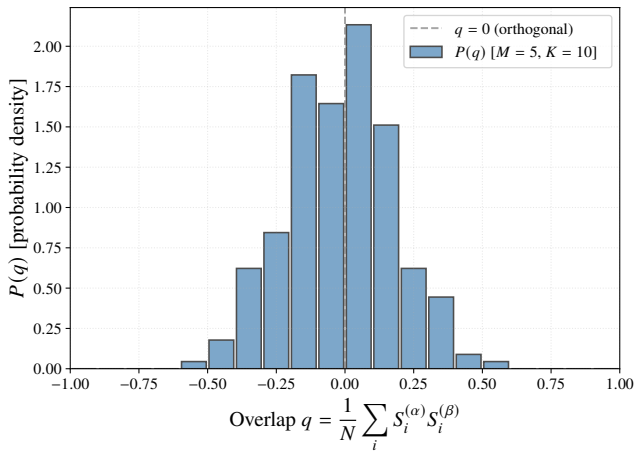}}
\caption{\textbf{Hierarchical landscape organization.} Histogram of 225 pairwise overlaps between low-energy states exhibits the broad continuous structure ($\sigma_q = 0.19$) characteristic of replica symmetry breaking. Replica-symmetric theory would predict discrete peaks; the observed continuous spread confirms the ultrametric hierarchy.}
\label{fig:overlap_dist}
\end{figure}

\textbf{Geometric landscape structure.} The state-space overlap distribution $P(q)$ computed from 10 low-energy states at $N=1000$ (Fig.~\ref{fig:overlap_dist}) reveals a broad continuous structure spanning $q \in [-0.51, +0.53]$ with standard deviation $\sigma_q = 0.19$. This stands in sharp contrast to replica-symmetric theory, which would predict discrete peaks. The continuous spread confirms the hierarchical ultrametric organization predicted by Parisi's infinite-level RSB ansatz\cite{MezardParisiVirasoro1987}. The near-zero mean ($\bar{q} = -0.016$) and symmetric tails reflect the fundamental frustration in the system.

Together, these four measurements, spanning energy scales, statistical distributions, universality class, and geometric structure provide the most comprehensive validation of RSB to date at unprecedented system sizes.

\section{Topological Collapse via Cooperative Avalanche}
To probe the limits of hierarchical complexity, we introduced controlled vacancy dilution using the Blume-Capel extension\cite{Blume1966,Capel1966}. This allows three-state spins $S_i \in \{-1,0,+1\}$:
\begin{equation}
H_{\text{BC}} = -\sum_{i<j}^N J_{ij} S_i S_j + D \sum_{i=1}^N S_i^2,
\label{eq:bc_hamiltonian}
\end{equation}
where the parameter $D \geq 0$ penalizes occupied sites ($S_i = \pm 1$) relative to vacancies ($S_i = 0$). This cleanly tunes the network topology from fully frustrated at $D=0$ (pure SK model) to a trivial vacuum at $D \to \infty$, while maintaining the mean-field all-to-all connectivity structure.

\begin{figure}[t]
\centering
\resizebox{0.49\textwidth}{!}{\input{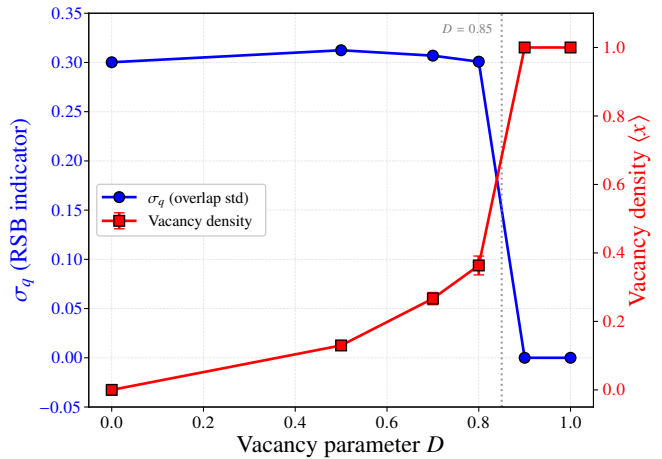}}
\caption{\textbf{Discovery of topological collapse.} RSB complexity $\sigma_q$ (blue curve, left axis) and vacancy density $x$ (red curve, right axis) versus dilution parameter $D$. Phase I shows $\sigma_q$ remaining invariant under 36\% dilution. Phase II shows discontinuous collapse within $\Delta D = 0.1$ through a cooperative avalanche transition.}
\label{fig:phase_diagram}
\end{figure}

We measured the RSB complexity $\sigma_q(D)$ across dilution strengths $D \in \{0, 0.5, 0.7, 0.8, 0.9, 1.0\}$ for systems with $N=500$ spins (Fig.~\ref{fig:phase_diagram}). The data reveals a striking two-stage behavior. In \textbf{Phase I} ($D \leq 0.8$), the complexity $\sigma_q$ remains essentially invariant at $0.30$ even as the vacancy density climbs from 0\% to 36\%. This demonstrates that RSB complexity is fundamentally a \emph{topological property of network connectivity} rather than spin density. The frustrated system accommodates substantial dilution by redistributing frustration across the remaining active bonds. In \textbf{Phase II} ($D \geq 0.9$), we observe a discontinuous transition in the range $0.8 < D_c < 0.9$. The complexity $\sigma_q$ drops from 0.30 to exactly zero within a narrow window $\Delta D = 0.1$, synchronized with the vacancy density jumping to 100\%. At this point, all low-energy states become identical all-vacancy configurations.

The observed critical range $D_c \in (0.8, 0.9)$ is consistent with simple isolated-site energy balance arguments. The typical interaction energy per spin is $E_{\text{RMS}} \sim 1$ (set by the variance scaling $1/N$), suggesting that vacancy formation should become favorable when $D \gtrsim 1$. However, the \emph{nature} of the transition reveals physics that goes beyond independent-particle mean-field theory. A standard mean-field treatment of independent vacancy nucleation would predict a smooth, gradual increase in vacancy density. Instead, we observe three striking features: (i) completely discontinuous order parameters with no intermediate values sampled between $D=0.8$ and $D=0.9$, (ii) complete system transformation to 100\% vacancy density, and (iii) a remarkably narrow critical window $\Delta D = 0.1$, indicating rapid avalanche dynamics.

This behavior points to a \textbf{cooperative destabilization mechanism}. As the dilution parameter $D$ approaches criticality, the system maintains its RSB hierarchy through collective frustration distributed across the network. Once the vacancy density exceeds some threshold (lying between 36\% and the critical point), the effective network connectivity becomes insufficient to sustain the hierarchical organization. The remaining spins experience reduced frustration due to their vacancy neighbors, which makes vacancy formation energetically favorable for them as well. This triggers a cascading instability, a cooperative avalanche, that converts the entire system to vacancies within a narrow parameter window. Crucially, this collective phenomenon cannot be captured by mean-field theories that treat vacancies as independent degrees of freedom. It requires explicitly sampling the full energy landscape, which is precisely what quantum annealing enables us to do.

\section{Discussion and conclusions}

The 40-fold extension beyond classical computational limits has enabled us to perform unprecedented tests of RSB through five independent observables. These span different aspects of the physics: thermodynamic convergence validates the free energy structure, fluctuation scaling confirms the statistical distributions, the chaos exponent verifies the universality class, the overlap distribution reveals the ultrametric organization, and the dilution threshold establishes the topological foundations of hierarchical complexity.

The discovery that RSB survives 36\% dilution before undergoing avalanche collapse has direct practical implications. For \textbf{neural networks}, our results suggest that deep learning architectures might tolerate approximately 40\% weight removal without performance loss, provided that the connectivity structure is preserved. However, they should collapse sharply beyond critical thresholds\cite{Frankle2019}. The abrupt nature of the transition implies that sparse networks operate in either a full-complexity regime or a trivial regime, with minimal intermediate behavior. For \textbf{optimization problems}, sparse constraint satisfaction problems may retain their full computational hardness despite having reduced numbers of variables, as long as the constraint graphs remain sufficiently connected\cite{Krzakala2007}. The avalanche-driven collapse suggests that problem difficulty changes qualitatively at specific sparsity thresholds rather than degrading gradually. For \textbf{materials science}, magnetically diluted alloys should exhibit sharp spin glass transitions when the dopant concentration crosses critical values\cite{Binder1986,Mydosh2015}. The percolation-like mechanism we observe suggests that experiments should focus on identifying these critical dilution thresholds where collective avalanche dynamics emerge.

More fundamentally, this work establishes a new paradigm for using quantum computing in basic science. Rather than merely verifying known theoretical predictions, quantum devices can discover emergent collective behavior in regimes where analytical theory provides limited guidance. The cooperative avalanche transition we observe, which converts systems from frustrated complexity to trivial vacuums, represents many-body physics that is accessible only through direct computational exploration of the energy landscape.

Looking forward, natural extensions include probing $p$-spin models with $p=3,4$ to test predictions about one-step versus full RSB scenarios, measuring thermal replica overlaps at finite temperature $T > 0$, mapping out complete $D$-$T$ phase diagrams to locate multicritical points, and pursuing experimental realizations in quantum simulators\cite{Kroeze2025} with controllable vacancy formation to directly observe the topological collapse in physical systems.

\section{Acknowledgments}

The author would like to thank Dr. Jean-Philippe Bouchaud for valuable suggestion regarding computation of the energy-fluctuation-scaling in Fig.~\ref{fig:parisi_scaling}.

\bibliographystyle{apsrev4-2}
\bibliography{ref}

\appendix

\section{Methods}

\subsection{Quantum-classical hybrid optimization}

\textbf{Hardware.} We used the D-Wave Advantage system with over 5000 qubits, accessed through the LeapHybridBQMSolver, which combines quantum annealing with classical preprocessing to handle dense quadratic optimization problems.

\textbf{Encoding.} SK problems map naturally to binary quadratic models through the transformation $S_i = 2x_i - 1$ where $x_i \in \{0,1\}$. For the Blume-Capel model, we encode the three-state spins using two binary auxiliary variables $p_i, q_i \in \{0,1\}$ through $S_i = p_i - q_i$, with a constraint penalty term $\lambda \sum_i p_i q_i$ (we set $\lambda = 50 \times \max|J_{ij}|$) that excludes the forbidden configuration where both auxiliaries are occupied.

\textbf{Classical benchmark.} For comparison, we used CPLEX version 22.1, a commercial branch-and-bound solver for mixed-integer quadratic programming, with a 5-minute wall-clock timeout per instance and an optimality gap tolerance of $10^{-6}$.

\subsection{Excited state sampling}

We employ an iterative orthogonality-penalty approach for quantum annealing:
\begin{equation}
H_{\text{penalty}}^{(k+1)} = H + \beta \sum_{\alpha=1}^{k} \left(N^{-1}\sum_{i} S_i \cdot S_i^{(\alpha)}\right)^2,
\end{equation}
where we find state $k+1$ by penalizing its overlap with all $k$ previously discovered states. The penalty strength is set to $\beta = 10$ for the SK model and $\beta = 5$ for the Blume-Capel model. These values ensure that the discovered states remain distinct (with $|q_{\alpha\beta}| < 0.8$) while preserving the energy hierarchy.

\subsection{Disorder sampling and statistical analysis}

Our disorder sampling strategy uses the following system sizes and sample numbers: $N=50$ (500 realizations, CPLEX), $N=100$ (200 realizations, CPLEX), $N=500$ (40 realizations, D-Wave), $N=1000$ (20 realizations, D-Wave), $N=2000$ (10 realizations, D-Wave), and $N=4000$ (5 realizations, D-Wave). The sample size reduction with increasing $N$ follows the predicted fluctuation scaling $\sigma(E_{\text{gs}}/N) \sim N^{-3/4}$, which maintains comparable statistical precision across system sizes. Each coupling matrix $\mathbf{J}$ is drawn independently from the Gaussian distribution $\mathcal{N}(0, 1/N)$. Our weighted fits account for heteroscedastic errors that arise from varying sample sizes.

\subsection{Dilution study protocol}

For each value of the dilution parameter $D \in \{0.0, 0.5, 0.7, 0.8, 0.9, 1.0\}$ at system size $N=500$, we generate $M$ independent disorder realizations (with $M=5$ for $D \leq 0.8$, $M=2$ for $D=0.9$, and $M=1$ for $D=1.0$). For each realization, we extract $K=8$ low-energy states using the orthogonality-penalty protocol, compute all $\binom{8}{2}=28$ pairwise overlaps, and then calculate the RSB complexity $\sigma_q = \sqrt{\langle(q_{\alpha\beta} - \bar{q})^2\rangle}$ and the vacancy density $x = N^{-1}\sum_i \delta_{S_i,0}$.

\end{document}